\documentclass{collective-intelligence}

\usepackage{graphicx}
\usepackage{url}

\begin{document}
\bibliographystyle{agsm}

\title{MARKERLESS MOTION CAPTURE IN THE CROWD}
%
%
%
%
%

\numberofauthors{1} 
%
\author{
%
%
\alignauthor
Ian Spiro, Thomas Huston, Christoph Bregler\\
       \affaddr{Department of Computer Science, Courant Institute}\\
       \affaddr{New York University}\\
       \email{ \{ian, thomas, chris\}@movement.nyu.edu}}

\maketitle
\begin{abstract}
This work uses crowdsourcing to obtain motion capture data from video recordings. The data is obtained by information workers who click repeatedly to indicate body configurations in the frames of a video, resulting in a model of 2D structure over time. We discuss techniques to optimize the tracking task and strategies for maximizing accuracy and efficiency. We show visualizations of a variety of motions captured with our pipeline then apply reconstruction techniques to derive 3D structure. \end{abstract}

\section{Introduction}

In traditional motion capture, a capture volume is established with several high speed cameras that track retroreflective dots on actors' clothes. This is acceptable for high-budget animation projects where entire scenes will be rendered from scratch with motion capture applied to animated characters. But the equipment is expensive and in many cases it  simply isn't possible to fit an activity of interest into a capture volume. There is increasing interest in motion capturing sports activities, either for use in games, movies, or sports medicine. Sports are notoriously difficult to motion capture, as they tend to require large capture volumes and the motion capture suits may interfere with the motions in question. A broader limitation of traditional motion capture is that it must be arranged for in advance. Many of the most interesting motions to study are found in history. We have vast archives of human motion recorded in video but without explicit information of the subject's joint configuration over time, as we get in motion capture.  Automatic video-based tracking without motion capture markers is a very active field in Computer Vision, but so far no working general solutions have been proposed (as we discuss in the Vision-based Tracking section).

With this motivation, we have built a system for capturing motion from video. We developed a tool for users to perform annotation of arbitrary sets of points in a video. The interface is deployable in a web browser so we can pay users of Amazon Mechanical Turk (MT) to complete the complex tracking task. With the aid of keyframe interpolation, a person can perform our task efficiently. We also deploy tasks redundantly to exploit the powerful effect of averaging in a crowd.

We have run the tool on several sets of data. The first study aimed to build a library of politicians' gestures and body language. The second study was of academic lecturers. This work, described in \cite{spiro2010hands}, approached the simplified motion capture problem of tracking a subject's head and hands over time.   After success with these tasks, we turned to something more complex: the annotation of high speed sports footage, in particular, baseball pitches. We increased our marker set from three points to thirteen and added a qualification test to further assure the quality of data.

The rest of this paper details the design of our user interface and our experiences deploying it on MT. We show examples of annotations, and assess the quality of the derived motion capture data including the effect of redundant HIT deployment. We then demonstrate several applications of our system.

\begin{figure}
\includegraphics[width=3.5in]{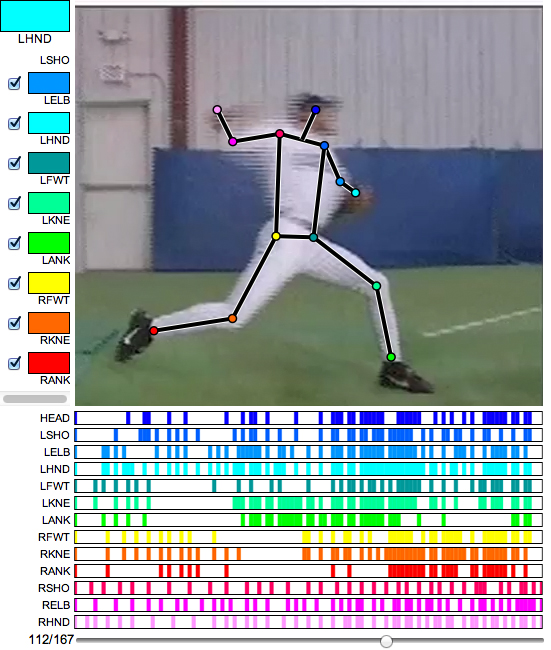}
\caption{\label{fig_ui}Our user interface for tracking, showing a completed 13-point baseball pitch annotation. Each marker and corresponding track is indicated with a different color. Tracks shows a line for every defined keyframe. }
\end{figure}

\section{Related Work}

\begin{figure}
\includegraphics[width=3.5in]{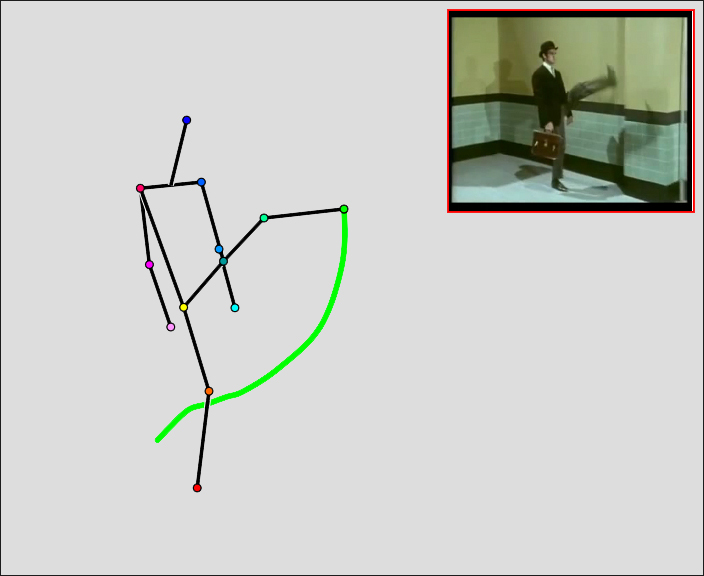}
\caption{\label{fig_ui}Motion visualization. Here we see a reconstruction of John Cleese's iconic Silly Walk.}
\end{figure}

\subsection{Annotation Tools}

The task of tracking the position of features or objects in a video, and ``matching'' the 3D pose and configuration of rigid and articulated objects in 3D is, in the visual effects industry, known as ``match-moving''  or ``rotoscoping.''  Methods date back over 100 years when animators traced film on light tables to produce animations.  Most recent high-end tools are based on supplying the user with an interface that is similar to 3D key-frame animation (such as that of Autodesk Maya and other tools).  The user can ``scrub'' the video back and forth, and can click on locations to set key-frames.  The frames in between are either interpolated or automatically tracked with general pattern trackers (\cite{AfterEffects,bojou} to name a few).  Some recent advances blend explicit notation and automatic tracking in an interactive way \cite{buchanan2006interactive,bregler2009ilm}.  
Another community, which includes gesture and multi-modal communication researchers, use a different set of annotation tools.  The most popular ones are ANVIL \cite{ANVIL} and MacVisSTA \cite{MacVisSTA}. Both tools are more targeted for time-based annotation of text tags, but have some capability of annotating spatial information. Neither tool permits annotation of 3D configuration information. 
All tools discussed so far have in common a high level of complexity that requires a user to undergo some training. In some cases the level of training needed to use these tools is extensive.  This is not possible for MT users: they need to understand how to use an annotation tool in a limited amount of time.  Another problem with the high-end tools is that they are generally not platform independent.  For MT we don't want to require that users have a specific operating system for installing a binary application. Therefore we employ a web-based interface written in Javascript.

LabelMe \cite{labelme} and the Web Annotation Toolkit \cite{sorokin_toolbox} provide web-based toolboxes for image annotations. And most recently a video extension \cite{yuen2009labelme} has been reported.  In \cite{sorokin2008utility}, they build a system that obtains coarse pose information for single frames but without video support. Another project \cite{vondrick2010efficiently} is specifically geared toward video of basketball but the resulting data is a set of human-size bounding boxes with no details of body pose. For our specific domain, these toolboxes do not provide the necessary functionality, since our annotations generally require handling of non-rigid objects and nuanced, high-speed motions.

\subsection{Vision-based Tracking}

\label{sec_visionbasedtracking}
General vision-based human body tracking has gained increased attention in recent years \cite{bregler1998tracking,deutscher2000articulated,felzenszwalb2005pictorial,sidenbladh2000stochastic,sminchisescu2001covariance,mori2002estimating,ramanan2005strike} but typically breaks on complex motions, motion blur, low-resolution imagery, noisy backgrounds, plus many other conditions that are present in most standard real-world videos.  It is beyond the scope of this paper to review all related tracking techniques and we refer to \cite{forsyth2005csh} for a survey.  Fully automatic tracking of gestures is not solved yet for the general case.  Part of our research agenda is to build a training database that will be used to develop and further improve such automated gesture tracking systems.

\begin{figure*}[!ht]
\includegraphics[width=7.2in]{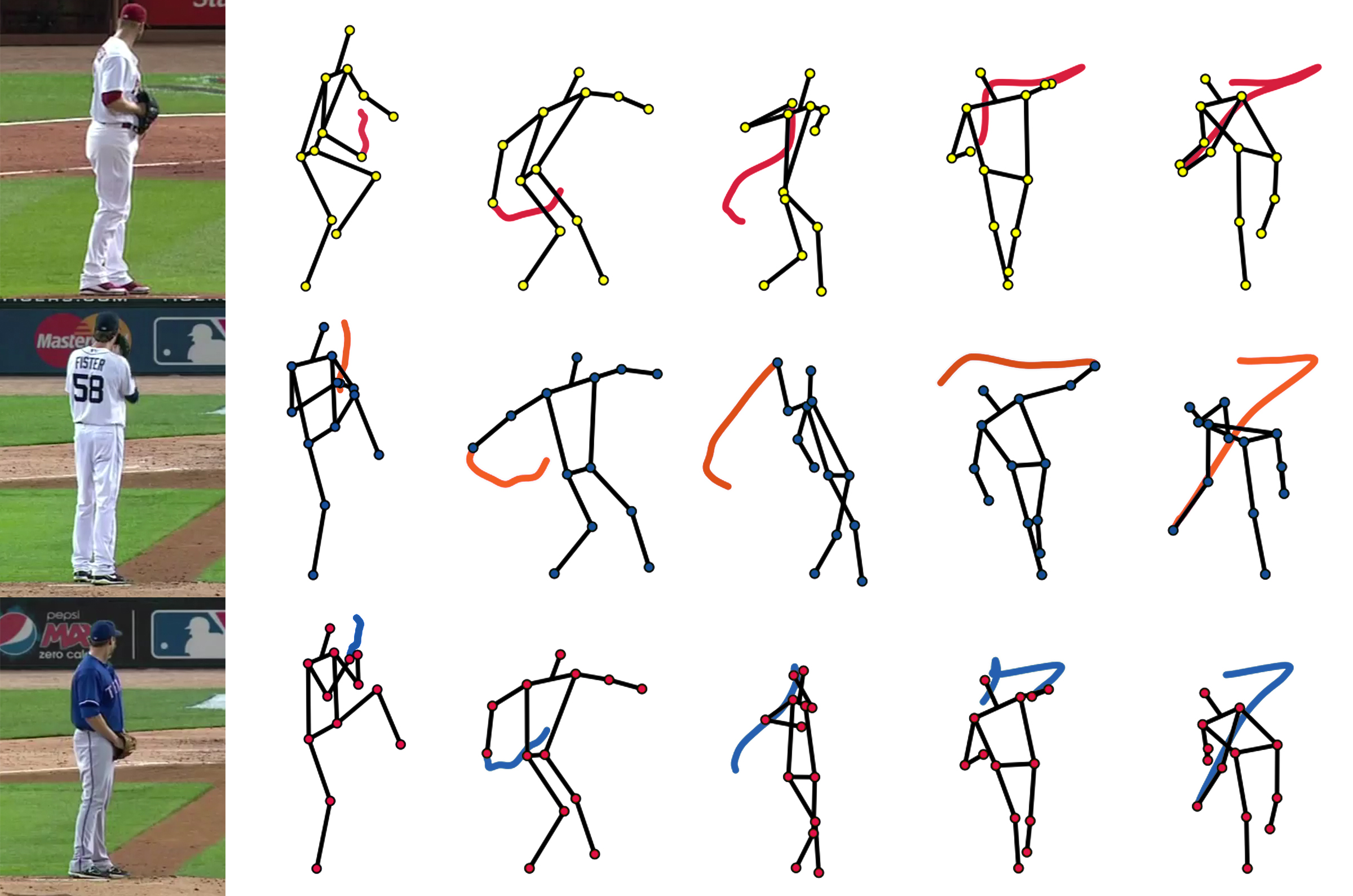}
\caption{\label{fig_pitches} Example motion summaries from our system. Each row is a different MLB pitcher (Chris Carpenter, Doug Fister, and Colby Lewis), shown at five different stages of a pitch (knee up, arm back, foot down, ball release, and follow through). The motion trail of the right hand is shown for frame t-10 through t. Note that while all three deliveries follow the same general pattern, each pitcher has a unique signature to their motion. In the knee up stage, while Fister and Lewis raise their left knees to almost the same level as their left hands, Carpenter's left knee never goes above his waist. Leading up to the ball release, Fister moves his right hand along a flat plane, while Carpenter and Lewis angle their arms more. We also see that Lewis has a distinctive twist of his right arm in between the foot down and ball release stages. And in the final stage, Carpenter's follow through is quite sharp, with his arm moving back along the same path as his ball release, whereas Fister's and Lewis's right arms form a v-shape between the ball release and follow through.
}
\end{figure*}

\section{Our Approach}

\subsection{User Interface}

The skill of motion tracking benefits from experience but is a task that most anyone should be able to understand and perform with minimal training. The task is potentially tedious in that it involves precise clicking of many points and scrubbing short snippets of video repeatedly. Annotating 10 points in a video of 100 frames potentially requires 1,000 clicks. However, a more efficient formulation is possible. In real video of human motion, particular points may not move at all during a particular range. In this case, by specifying the frame and location when the point stops moving $(x_1,y_1,t_1)$ and the time $t_2$ when the point begins to move again will compactly define the location of a point in multiple frames. Another possibility is that a point moves linearly between two locations over a range of frames. Here two $(x,y,t)$ designations can convey many frames worth of annotation. For the sake of decomposing labor, this means it is better for a single individual to annotate a specific point over a range of frames, as opposed to many points for one frame.

The interface shown in \ref{fig_ui} was built in Javascript and takes advantage of HTML5 for playing video files. A user is typically asked to track 3-4 points over 100-200 frames using it. Each part appears as a clickable colored square as in a paint program. When the user clicks on the video panel, a marker of the selected track appears. A line is also made underneath on the appropriate track visualization to indicate a keyframe has been placed at the current time index. This helps to remind the user of the keyframe process. Some users get into trouble when they place a frame incorrectly, then scrub back near the frame and try to fix it by adding additional keyframes. If the original incorrect keyframe isn't removed or fixed, the motion will always have a problem. Users can change the position of previously placed markers simply by dragging. Users can jump to a particular track and keyframe by clicking the lines on a track visualization.

A key aspect of the interface is the association of keystrokes with all major actions. This feature greatly speeds up the tracking task since the user can keep his cursor near the action of the video without moving it away to click a button. This is a natural consequence of Fitt's Law  \cite{fitts1954information}, which generally says the time it takes a user to click a given point grows logarithmically with the ratio of target distance over target size. In this task, click accuracy and efficiency are of great importance, affecting either the quality of our results or the effective pay necessary on MT. 

The process of tracking a motion is iterative. The user can make a first pass using the forward button to skip 5 frames, create a keyframe, and repeat for a coarse annotation. Then the user plays back the motion to see how the tracking looks. If more keyframes are necessary, as is common in time ranges with rapid or non-linear motion, the user adds additional keyframes. Our interface also supports zooming, so the user may more precisely indicate markers.

We indicate the points to track with 4 letter symbols, like HEAD, LANK, RELB, which correspond to typical marker sets used in motion capture. We include a diagram with each part marked on a person and display a textual description to go along with the selected track. An important detail with motion tracking is to prevent the user from annotating with a left-right flip. If the video contains a subject facing out, then body parts are mirrored relative to the user. You can define left to be the user's left or the subject's left, but either way some workers may get it wrong. In our heads and hands interface, we devised the simple solution of popping up a large warning that appears anytime the user seems to have it backwards. For baseball annotations, we describe the LHND point as "the hand wearing the glove" and the RHND point as "the throwing hand," and this effectively prevented flipping. (Note that we are only studying right-handed pitchers for now.)

\subsection{Qualification Task}

In our first attempts at markerless motion capture on MT, we simply checked all completed assignments and determined which to approve or reject. This took up a large amount of researcher time and would not scale. Next we attempted to deploy HITs redundantly with multiple assignments and use the median to improve accuracy. This required more total work for MT and we still were not satisfied with the general quality of results. Many workers would annotate a single frame and submit the assignment while others clicked randomly or otherwise achieved very low accuracy. In the end, we still had to manually check most of the work. 

We now require all Turks who wish to complete our motion capture HITs to pass a qualification test. We use the exact task that is demonstrated in a video that plays during HIT preview. The user must track 3 points over 90 frames of a baseball pitcher throwing a ball. Our test is scored automatically using as an answer key, a completed annotation by a tracking expert in our research group. A user attempting the qualification test is allowed to request a score at any time, which is computed automatically by comparing all 270 data points to our answer key. The result is a sum distance which is then scaled to a score out of 100\%. We scale the test such that a score of 80\% is satisfactory but not perfect. A user attempting the qualification test is only allowed to submit their work after they have obtained 80\% or higher and we record the final score. By allowing some range of quality, we have created a kind of honeypot to mark Turks who may be less motivated. For some HITs, we can accept our minimum 80\% quality score while for others, we want the most talented and motivated workers. One recommendation we have is to pay users to take qualification tests instead of making them unpaid, like most on MT. This helps to increase the labor pool and with the auto-scoring system, you only to pay for satisfactory completions.

\begin{figure}
\includegraphics[width=3.5in]{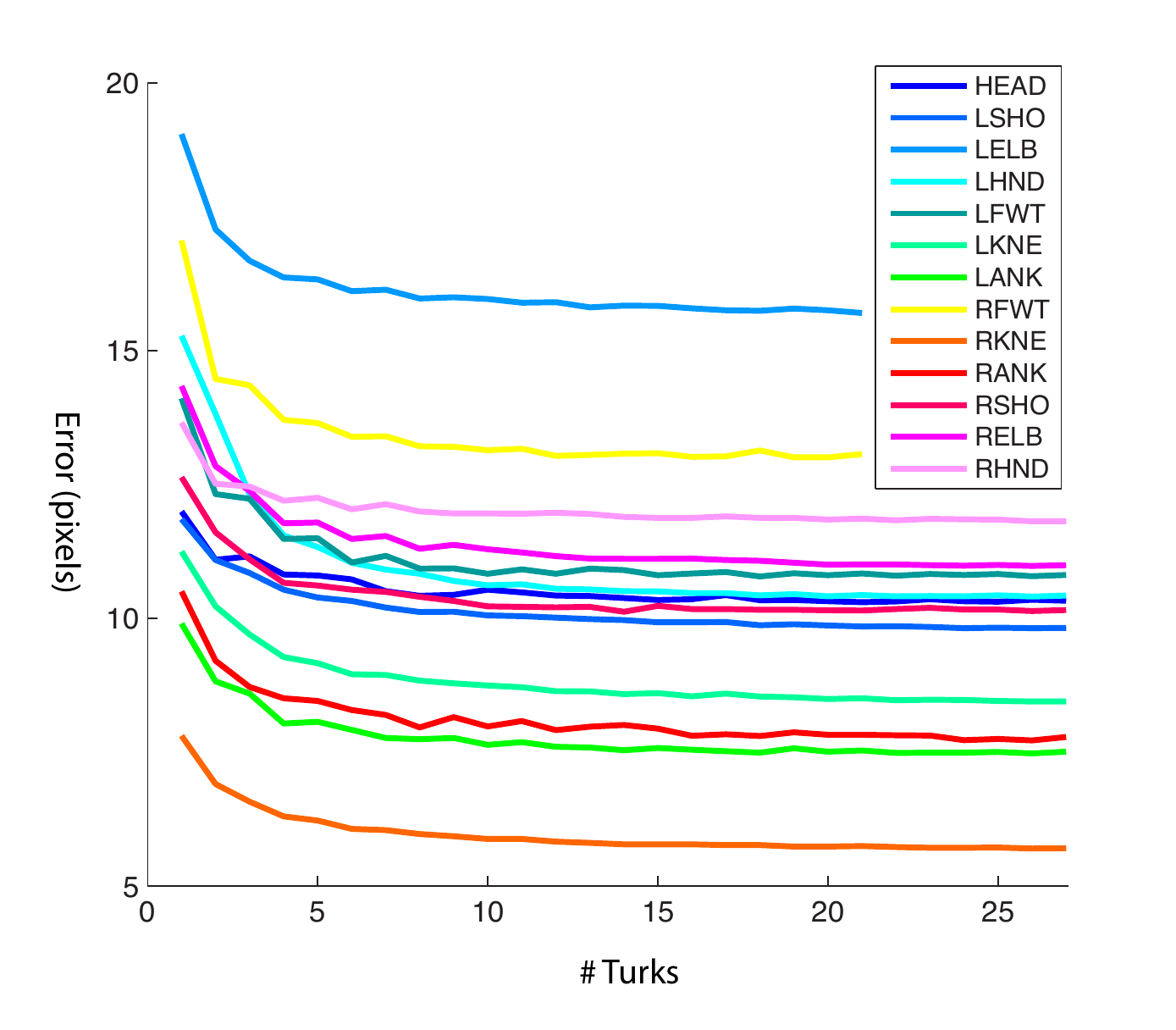}
\caption{\label{fig_ox}The median error for tracking 13 points, measured in pixels, shown for different crowd sizes.}
\end{figure}

\subsection{Cost}

Using the previously described interface and qualification system, we are able to annotate complex footage of human motion at a price considerably lower than what a studio pays for comparable tracking services. We typically create HITs that involve annotating 3 tracks over 100 frames which takes 5 to 10 minutes when done efficiently. We pay 50 cents per hit, resulting in an hourly wage of \$3-\$6. Table 1 shows our effective cost to annotate pitches. This is the cost to get 13 markers of one pitch tracked by one Turk each. We typically deploy each HIT 5 times and use the median of each marker location to boost the quality of our tracking. Of course, this costs 5 times as much.

\begin{table}
\centering
\caption{HIT Details} Single Turk per track
\begin{tabular}{|c|c|c|c|c|c|} \hline
Frames&Tracks&HITs&\$/HIT&\$/pitch&\$/min\\ \hline
100 &13 & 4& \$0.50 & \$2 & \$36 \\ \hline
\end{tabular}

\end{table}

\section{Results}

\subsection{Accuracy}

To assess the accuracy of our human-powered markerless motion capture system, we used traditional motion capture data as ground truth. We recorded a baseball pitcher in a motion capture studio along with video data, then performed camera calibration so we could project the 3D data to align with the video in each frame. We divided the tracking task across 4 HITs, each one 115 frames with 3-4 tracks. We deployed each HIT to approximately 30 MT workers. 

We denote the true location of a particular marker:
 \[
P = \begin{bmatrix}
       x           \\[0.3em]
       y 	\\[0.3em]
     \end{bmatrix}
\]

The location of a marker at time  \begin{math} t \end{math} is  \begin{math} P_t  \end{math}. We define \begin{math} \hat P_t \end{math} as an estimate of a marker's position and compute it by taking the median of any number of user estimates, as obtained from our annotation tool. We define pixel error for any marker at a time frame  as the L2 norm of the difference between the estimate and ground truth:

\begin{center}
\begin{math} E^{Pixel}_t =  |P_t -  \hat P_t | \end{math} \\
\end{center}

Thus pixel error for a given marker is simply the per frame error averaged over all frames. Table 2 shows a summary of the results. All numbers are reported in pixel values and the input video was 640x480. For reference, 10 pixels is approximately 5 cm in the real world given our particular camera configuration. There is a considerable amount of error and it varies across the marker set. The knee and ankle seem to be the easiest to track and this could be because the lower markers move slower than the upper ones in a baseball pitch and also because this area of the video had higher contrast. In the case of the hand and elbow markers, there were many frames where it was virtually impossible to distinguish the pitcher's hands from his torso. This is a limitation of our recorded video. (On the other hand, the low contrast meant that it was nearly impossible to see the actual retroreflective markers in the video, which could have made the task too easy or less representative of markerless motion capture.) Further examination of the errors over time showed that Turks had trouble consistently locating subtle 3D configurations in 2D. For example, they were asked to mark the head as the point on the surface of the bridge of the nose between the eyes, but the majority of Turks clicked the center of a circle that approximately bounds the head. This means they were actually tracking a point in the center of the skull. As an alternative assessment, we propose a measure of the motion itself, denoted in the table as Motion Error. We define the motion for a marker at time t as the vector:

\begin{center}
\begin{math} M_t = P_t - P_{t-1} \end{math} \\
\end{center}

We compute motion error for a particular frame and marker as:

\begin{center}
\begin{math} E^{Motion}_t =  |M_t -  \hat M_t | \end{math} \\
\end{center}

Again, we average across all frames and report the errors in the last two columns of Table 2. Using this measure, a Turk may have the wrong conception of a marker location, but if he tracks the wrong location correctly, motion error should not be affected. Note that pixel error and motion error cannot be directly compared since motion error is a velocity measured in pixels/frame. We can still look at the relative errors between different body parts to see, for example, that Turks get a similar amount of absolute pixel error  when tracking shoulders and hips, as compared to hands and elbows. But when we look at the motion error, we see that the Turks do quite a bit better with tracking the motion of shoulder and hips. This makes sense because the shoulders and hips move slower than elbows and hands.

Despite the reported error, the animations of markerless motion capture seem to  capture some large gestalt of a subject's motion. The Turks are relying on their own perceptual faculties to perform tracking, which may not be ideal for absolute tracking. A person viewing the resulting animation has a perceptual system comparable to that of the original Turk, which could explain why, subjectively, visualization results look good.

\begin{table}
\centering
\caption{2D Accuracy of Motion Capture}
\begin{tabular}{| r | c | c | c | c  | } \hline
& \multicolumn{2}{|c|}{Pixel Error}  &\multicolumn{2}{|c|}{Motion Error}  \\ \hline
Part & L & R & L & R \\ \hline
Shoulder& 9.92& 10.15& 1.96 & 1.89 \\ \hline
Elbow& 14.97& 11.02 & 2.87 & 4.03\\ \hline
Hand& 10.35& 11.28 & 3.26 & 4.26\\ \hline
Hip& 10.57& 11.18 & 1.43 & 1.56\\ \hline
Knee& 8.54& 5.37& 1.82 & 1.30\\ \hline
Ankle& 7.14& 6.42& 1.26 & 1.31\\ \hline
Head& \multicolumn{2}{|c|}{8.93}  & \multicolumn{2}{|c|}{1.24}\\ \hline

\end{tabular}
\end{table}

\begin{figure*}[!ht]
\includegraphics[width=7.2in]{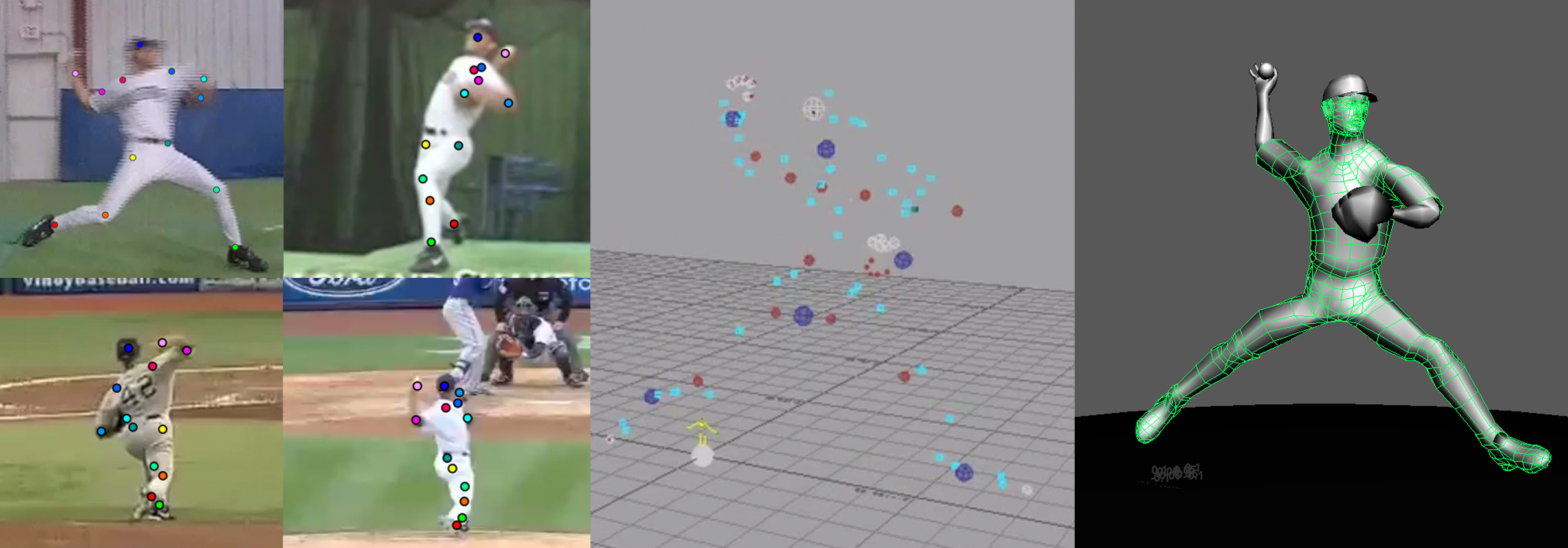}
\caption{\label{fig_pipeline} Phases of 3D reconstruction. We deployed several videos of baseball pitcher Mariano Rivera through the annotation pipeline. Factorization-based reconstruction produces an approximate skeleton animation. This data is cleaned up by an artist and applied to a character model to produce a 3D animation.
}
\end{figure*}

\subsection{Crowd Effects}

We are hoping to take advantage of the power of the median to provide the most robust estimates for our tracking data. The Wisdom of Crowds \cite{wisdomcrowd} describes many historical examples of crowd success and a few failures while \cite{hortondot} studies this explicitly with a guessing game and reveals that crowds are not immune to systematic bias. To study this effect as it relates to motion tracking, we deployed certain baseball annotation HITs to many Turks and studied the change in error as we included larger sets of Turks in the crowd. We selected random subsets of the 27 Turks, ranging in size from 1 to 27. For each subset, we take the median tracking result and compute the error of every point relative to ground truth, as described in the Accuracy section. We repeat many times with different permutations to reduce noise. We take the mean of all of these errors and report the results in Figure \ref{fig_ox}.  We note that error decreases for all markers as we increase the size of the crowd but that most of this averaging advantage is realized with 5 workers. Applying a linear regression to each marker line in the range from 10 to 27 results in slopes around -.01. This means, assuming the linear relationship holds, it will take 100 additional Turks for each desired additional pixel of accuracy. This appears to be prohibitively expensive, so more research needs to be done to improve tracking tools or the instructional design of the HIT.\newline\newline

\subsection{Visualization and Search}

A result of our effort is the ability to visualize motions. One possible visualization is video playback of the annotation as a stick figure. We can compare stick figures side by side for an analysis that focuses on the motion instead of pixel-level details. We can overlay a trace of any marker over time to emphasize and convey the shape of that marker's motion. (See our supplemental video for examples of this technique, located at http://movement.nyu.edu/markerless.) The second option is to create static images that summarize a motion by concatenating several frames along with marker traces as in  figure \ref{fig_pitches}. With this technique we can illustrate a complex motion with a single, static image.

\begin{figure}[b]
\centering
\includegraphics[width=\columnwidth]{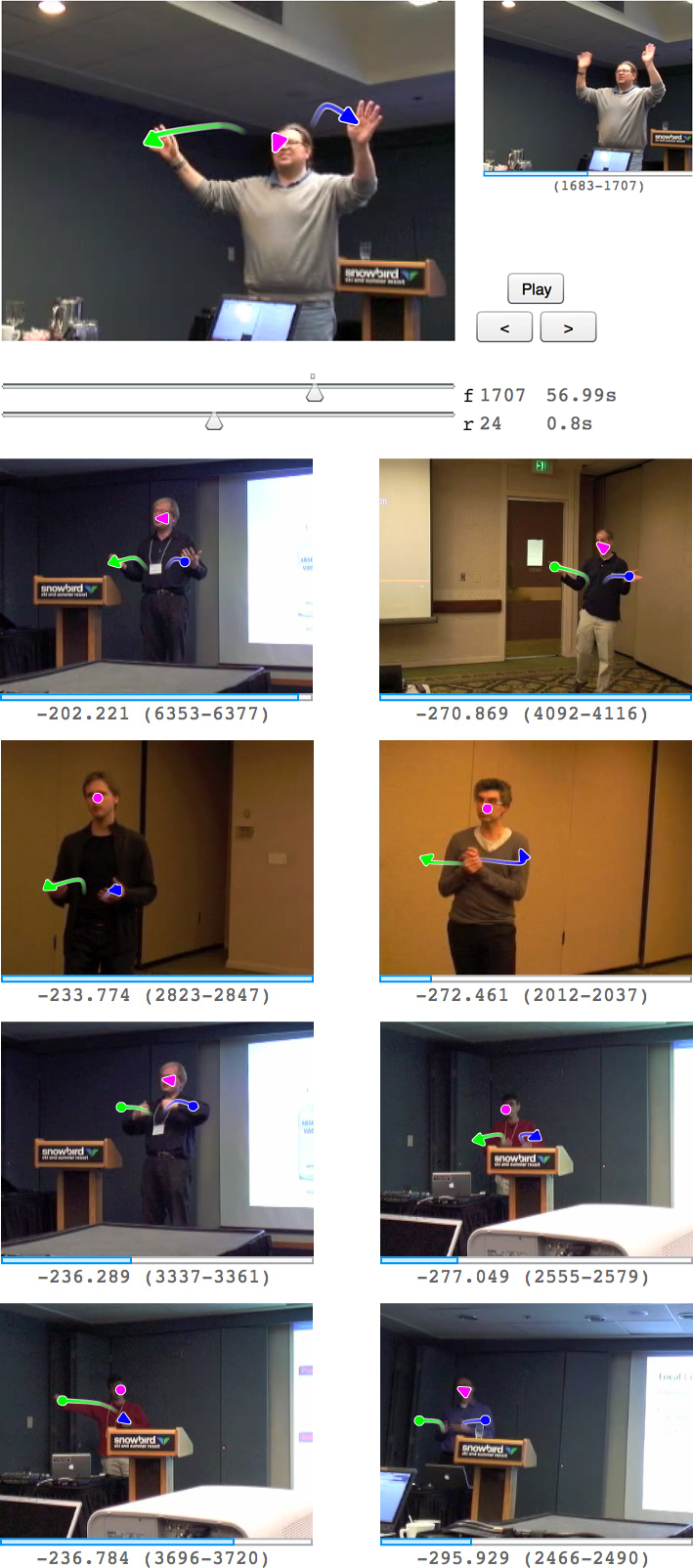}
\caption{\label{fig_gesture_match} Gesture search by example. The user scrubbed through the top video until a suitable handwaving gesture was found then used this as the search input. The search returned the 8 closest matches.}
\end{figure}

Another result of our system is the ability to search for gestures by example. We used the MT-based annotation pipeline on a set of academic videos from a recent conference at Snowbird in Utah.  Similar to a search engine on text, or the new Search by Image feature on Google, we can build a fast search engine that is motion or gesture based.  We store all videos that are annotated with our MT pipeline in a database and index the database entries with a motion-based hashing technique that uses the head and hand motion vectors in a window of frames. The hashing scheme is designed to be robust to changes in global position or scale variations. Figure \ref{fig_gesture_match} shows an example of inputing a video with a hand-waving gestures and we get back a ranked list of videos that have similar motions.

Another application of this crowd-facilitated gesture annotation can be used for Òbody language analysisÓ of speakers.  In previous research \cite{Williams2008,Williams2010} we developed an analytical technique that can summarize what most frequent gestures for a specific speaker, or how two speakers compare to each other, i.e. how many idiosyncratic 
gestures they have in common.

\subsection{Machine Learning}

We have started to employ the annotation pipeline to quickly build corpora of
annotated examples for training learning-based systems.  For instance, we used the dataset from the Snowbird
conference of academics' head and hand locations to train an automatic pose estimation system
\cite{taylorpose11}.  Using a deep network training architecture, the
system was able to automatically annotate human body poses
in visually challenging environments and was able to outperform
many other state-of-the-art pose estimation techniques reported in the
computer vision community. 


\subsection{3D Reconstruction}

Until this point, we've only discussed our system as a way to recover 2D motions. In general, recovering 3D from a single 2D image is ill-posed, especially in cases with complex, non-rigid motions. However, when multiple camera views of the same scene are available, it may be possible. We took archival footage of Mariano Rivera, a well-known baseball player with a distinct pitching style, that showed him from several different angles and ran each view individually through the pipeline. We applied a factorization-based 3D reconstruction technique (related to \cite{torresani2008nonrigid}) that can reconstruct the body markers with high accuracy in 3D.  This was then used to transfer to a 3D animation package and used for some baseball visualizations for a project with the New York Times \cite{nytnyu}. The system produces skeleton motion of the baseball pitcher including subtle, high-speed dynamics. With some artist cleanup, the final result is a compelling 3D animated re-creation of a human motion that was never seen in a motion capture studio.

\section{Discussion}

This work demonstrates the possibility of a markerless motion capture pipeline powered by human intelligence. With a keyframe-based user interface, we make a task that can be performed cheaply by a general pool of information workers with a resulting signal that can be boosted through redundant task deployment. We note that our system is limited in terms of absolute pixel accuracy but nonetheless, creates nuanced visualizations of motion. This can be observed in the various animations we have produced of natural human gesture, stylized walks, and sport motions. Subjectively, these animations capture some essence of a subject's motion including subtle dynamics like bounce. Our use of a complex, auto-scored qualification test allows us to simultaneously train workers and assess their ability. We believe that a harder qualification test requiring more exact positioning of markers could result in a better labor pool and better absolute pixel accuracy for future work. We also plan to develop an interface for Turks to smooth jittery motions by applying bezier-based smoothing and manual adjustment of control points.

We show several applications of our 2D annotations including static visualization and animated playback. We show that 2D tracking of just 3 markers is enough to facilitate example-based search across a corpus of data or to create single-subject motion profiles by clustering. Finally, we demonstrate that several 2D annotations of one subject from different camera angles can be used to create a viable 3D reconstruction even with a motion as nuanced as a baseball pitch.

The pipeline is invaluable for our work in human motion analysis. It opens up the possibility of capturing motion from historical video archives or for future recordings in scenarios that are incompatible with a traditional motion capture studio.

\section{Acknowledgements}
We would like to thank Graham Taylor and George Williams for their help and the Office Of Naval Research (ONR N000140910789, ONR N000140910076), Autodesk, and Google for supporting this research.

\bibliography{collective-intelligence}  
%

\end{document}